\documentclass[twocolumn,secnumarabic,amssymb, aps, prl, figure, superscriptaddress,floatfix, tightenlines] {revtex4-1}
\usepackage{graphicx}

\usepackage{braket}
\usepackage{multirow}
\usepackage{gensymb}
\usepackage{booktabs}

\begin{document}

\title{Rashba-Dresselhaus Effect in Inorganic/Organic Lead Iodide Perovskite Interfaces}%
\author{Chang Woo Myung}
\affiliation{Center for Superfunctional Materials, Department of Chemistry, Ulsan National Institute of Science and Technology (UNIST), Ulsan 44919, Korea}

\author{Saqib Javaid} 
\affiliation{Center for Superfunctional Materials, Department of Chemistry, Ulsan National Institute of Science and Technology (UNIST), Ulsan 44919, Korea}

\author{Kwang S. Kim}
\affiliation{Center for Superfunctional Materials, Department of Chemistry, Ulsan National Institute of Science and Technology (UNIST), Ulsan 44919, Korea}

\author{Geunsik Lee}
\affiliation{Department of Chemistry, Ulsan National Institute of Science and Technology (UNIST), Ulsan 44919, Korea}

\email{gslee@unist.ac.kr, kimks@unist.ac.kr}
\date{\today}%


\begin{abstract}
Despite the imperative importance in solar-cell efficiency, the intriguing phenomena at the interface between perovskite solar-cell and adjacent carrier transfer layers are hardly uncovered. Here we show that PbI$_2$/AI-terminated lead-iodide-perovskite (APbI$_3$; A=Cs$^+$/ methylammonium(MA)) interfaced with the charge transport medium of graphene or TiO2 exhibits the sizable/robust Rashba-Dresselhaus (RD) effect using density-functional-theory and ab initio molecular dynamics (AIMD) simulations above cubic-phase temperature. At the PbI$_2$-terminated graphene/CsPbI3(001) interface, ferroelectric distortion towards graphene facilitates an inversion breaking field. At the MAI-terminated TiO$_2$/MAPbI$_3$(001) interface, the enrooted alignment of MA$^+$ towards TiO$_2$ by short-strong hydrogen-bonding and the concomitant PbI$_3$ distortion preserve the RD interactions even above 330 K. The robust RD effect at the interface even at high temperatures, unlike in bulk, changes the direct-type band to the indirect to suppress recombination of electron and hole, thereby letting these accumulated carriers overcome the potential barrier between perovskite and charge transfer materials, which promotes the solar-cell efficiency.
\end{abstract}

\maketitle

Solar energy is a highly efficient and eco-friendly energy source for future energy harvesting. In recent years, inorganic/organic hybrid halide perovskite solar cell (PSC) based on ABX3 (A = Cs$^+$, CH$_3$NH$_3$$^+$ (MA$^+$), CHN$_2$H$_4$$^+$ (FA$^+$); B = Pb$_2$$^+$; X = Cl$^-$, Br$^-$ or I$^-$) have shown rapid progress achieving over 22 $\%$\cite{1} solar cell efficiency which is considered to be most promising large-scale solar energy materials.\cite{2}  PSC owns many interesting physical properties including giant dielectric screening,\cite{3}  bottleneck of hot phonon relaxation process,\cite{4}  multi-excitonic states,\cite{5}  and polaron state.\cite{6}  Owing to large electron-phonon coupling nature of PSC, Frohlich polaron state has been proved experimentally,\cite{7,8}  with its polaron radius being $\sim$ 4 unit cells.\cite{9}  This quasiparticle state would explain its good carrier transport property even in the presence of impurities.\cite{10}  Upon photoexcitation, electronic dielectric screening leaps orders of magnitude which help the exciton be dissociated with small binding energy.\cite{3,11}  Impact ionization of hot exciton with carriers is expected to be very high in perovskite nanocrystals giving rise to multi-exciton emission.\cite{12}  These interesting characters of PSC materials are ideal for practical solar cells and light emitting diodes.\cite{13}  Nevertheless, despite explosive discoveries in experiments, theoretical understandings underneath ongoing experiments are hardly made yet particularly regarding the Rashba-Dresselhaus (RD) effect at the interface between PSC and adjacent carrier transfer layers.
	The spin-orbit coupling (SOC) field, which is odd-in-k (momentum) and time reversal symmetric, in non-centrosymmetric crystals or at the interface of heterostrucures, gives rise to intriguing Rashba-Dresselhaus (RD) splitting. The effective low order perturbation terms of RD interactions are derived according to a given symmetry of the model. The lowest order Hamiltonian  in $kp$ is \cite{14}

\begin{widetext}
\begin{eqnarray}
H_{RD}(k) = \alpha_R (k_x \sigma_y + k_y \sigma_x ) + \alpha_D (k_x \sigma_x-k_y \sigma_y ), \alpha_{RD}= \frac{\Delta E}{2\Delta k}, 	(1)
\end{eqnarray}
\end{widetext}

where k is momentum, $\sigma_{i= x, y, z}$ is the spin Pauli matrices and the strength of RD interactions is defined by coupling constant $\alpha_{RD}$. The RD interactions are universal so that many systems such as noncentrosymmetric crystals, heterojunction,\cite{15}  metal surface,\cite{16}  and graphene\cite{17} show a sizable energy splitting. $\alpha_{RD}$ varies depending on systems ranging from 0.067 $eV \cdot \AA$ (InAlAs/InGaAs)\cite{18}  to 4.0 $eV \cdot \AA$ (Bi$_2$Se$_3$).\cite{19}  Recently, it is realized that perovskite solar cell materials containing heavy elements like Pb or I show large RD coupling constants: $\alpha_{RD}$ ~ 1.6 $eV \cdot \AA$ in 2D PSC (C$_6$H$_5$C$_2$H$_4$NH$_3$)$_2$PbI$_4$\cite{20}  and $\alpha_{RD}$ ~ 2.75-3.75 $eV \cdot \AA$ (in the original paper $\alpha_{RD}$ is 7-11 with different definition of $\alpha_{RD} = \frac{2\Delta E}{\Delta k}$)) in MAPbBr$_3$\cite{21}  and 3D CsPbBr$_3$ nanocrystal.\cite{22}  Previous studies on RD splitting in PSCs have focused on an inversion symmetry breaking in bulk phases with an artificial condition such as uniaxial pressure to trigger ferroelectricity in PSC.\cite{23}  Recent work clarified the importance of the RD effect on 1s exciton state of PSC.\cite{24}

Meanwhile, an interesting aspect has been realized that dynamical Rashba splitting occurs in both centrosymmetric I4/mcm and non-centrosymmetric I4cm tetragonal phases simulated by Car-Parrinello molecular dynamics.\cite{25,26} It has been proposed that on a large scale (> 8 $nm^3$) where the entropy of MA$^+$’s orientations is high, the RD effect might be quenched.\cite{26} An application to the spin filter device that makes the spin precess during the propagation in PSC has been proposed using RD interaction.\cite{27}  A technological impact is that the RD interaction changes the direct-type band structure to the indirect one to suppress the recombination of carriers and to promote carrier accumulation at the barrier between PSC and charge transfer materials. Although the understanding of interface phenomena in solar cell device is crucial, until now there is no work related to RD interaction at the interface of PSC and other material layers including electron and hole transport materials and the impact of RD interaction on solar cell performance. Graphene is a fascinating material for various applications such as transistors, optoelectronics, nanoelectronics, medical application etc.\cite{28}  Particularly multi-layer graphene has been proposed as an effective hole transfer material by its lower work functions close to the valence band maximum (VBM) of PSC.\cite{29}  TiO$_2$ is widely used for electron transfer materials (ETM) because of its transparency, ideal band alignment and synergetic effect with PSC.\cite{30,31,32} However, the RD effect at the interface between PSC and hole/electron transfer materials has not been studied yet.

In this work, for the first time, we clarify an elusive aspect of PSC heterostructures using the first principles calculations and AIMD simulations accounting for the RD effect. We have carried out the calculations for graphene/cubic-CsPbI$_3$(001) as a prospect interface for improving carrier transport and TiO2/cubic-MAPbI$_3$(001) as a well-known electron transport layer for PSCs device. Although we observed RD interactions at both interfaces, their mechanisms are different in intriguing ways. At PbI$_2$-terminated Gr/CsPbI$_3$(001), the ferroelectric Pb-I distortion promotes significant Rashba interactions both at 0 K and above 600 K. On the other hand, at MAI-terminated TiO$_2$/MAPbI$_3$(001), the direction of organic MA$^+$ near-fixed by strong short hydrogen bonding (SSHB),\cite{33,34}  even above 330 K and the concomitant distortion of PbI3 sublattice promote the RD interactions. Here, we show that unlike bulk where high entropic disorder of MA would quench the RD effect, the interfacial RD effect is robust in thermal effects and is beneficial for solar cell efficiency.

\begin{figure}
\centering
\includegraphics[width=8.6cm]{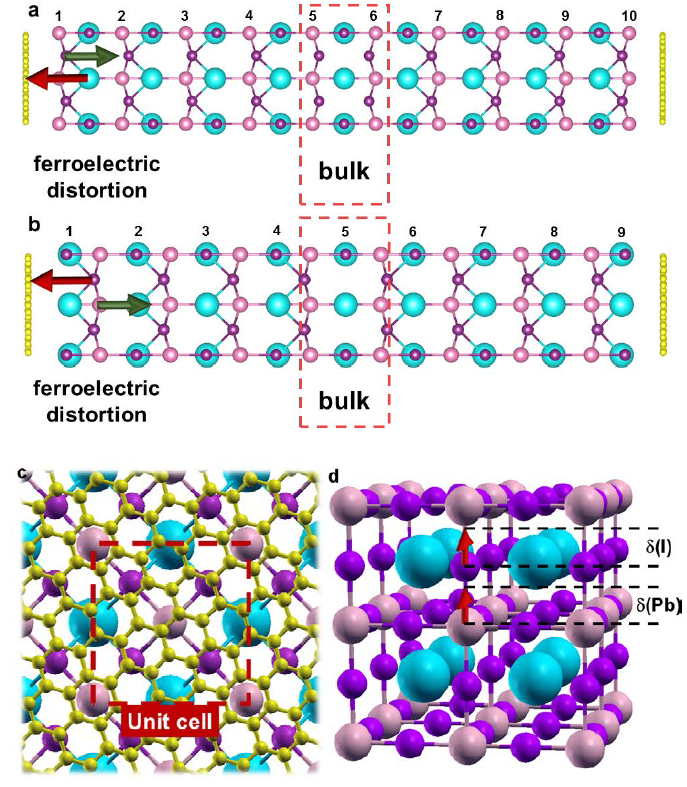}
\caption{\label{Figure. 1} (a) 10 layers of $(\sqrt(2) \times \sqrt(2) \times 1)$ (a) PbI$_2$-terminated and (b) CsI-terminated cubic CsPbI$_3$ for (001) surface sandwiched by graphene (yellow) viewed from [100] after structural relaxation. The number denotes the layer number. Each PbI2 layer undergoes a ferroelectric distortion of Pb (red arrow) and I (green arrow) along [001] direction with its magnitude gradually increasing towards graphene. For the correct description of the system, the bulk region (dashed red box) retained to have the inversion symmetry by surrounding it with 5-6 layers. (c) A unit cell of Gr/CsPbI$_3$(001)-$(\sqrt(2) \times \sqrt(2) \times 1)$ viewed from [001]. (d) Schematic displacements of Pb and I atoms near surface along [001] direction, denoted as $\delta(Pb)$ and $\delta$(I), respectively.}
\end{figure}

We constructed an interface of graphene/CsPbI$_3$(001)/graphene modeled by a slab of 10(9) layers of $(\sqrt(2) \times \sqrt(2) \times 1)$ PbI$_2$-(CsI-)terminated cubic CsPbI$_3$ with the lattice mismatch ~ 1.93 $\%$ between two systems (Fig. 1a, 1b and 1c). We confirmed that the surface dipole does not affect both the relaxed geometry and the corresponding electronic structures because the cation and anions (Pb$_2^+$I$^‒_2$ or Cs$^+$I$^‒$) at the termination are oriented parallel to surface. We adopt a symmetric slab sandwiched by graphene at each end to avoid any unphysical artifact. As reported from a previous LDA+D2 calculation of Gr/tetragonal-MAPbI$_3$,\cite{35}  we observe a ferroelectric distortion driven by an attraction between graphene and cation Pb$^{2+}$ in the PbI$_2$-termination. A measured distance between PbI$_2$ layer and graphene (Fig. 1a) is $d_{PBE-D3}$ ~ 3.21 $\AA$ at the PBE+D3 level and $d_{PBE-TS}$ ~ 3.28 $\AA$ at the PBE+TS level of theory which is smaller than LDA+D2 $d_{LDA-D2}$ ~ 3.45 $\AA$. As for CsI-termination, a distance between CsI layer and graphene (Fig. 1b) is $d_{PBE-TS}$ ~ 3.4 $\AA$ slightly larger than PbI$_2$-terminated surface. Despite the ferroelectric distortion, $C4v$ point symmetry of cubic structure is conserved which is manifested as pure Rashba type splitting in contrast to TiO$_2$/MAPbI$_3$ interface that will be discussed later. Atomic displacements near surface in both cases (Fig. 1a, b) are only significant along [001] (Fig. 1d). However, we observe that the displacement directions of both cases are opposite to each other. For PbI$_2$-termination, the displacements are $\delta$(Pb) ~ +0.4 $\AA$ and $\delta$(I) ~ -0.6 $\AA$, while for MAI-termination, the displacements are $\delta$(Pb) ~ -0.5 $\AA$ and $\delta$(I) ~ +0.6 $\AA$ (Fig. 1d). The ferroelectric displacement gradually increases when approaching the interface with its maximum at the very interface. Bulk-like (or inversion symmetric) layers, 5 th and 6 th layer (Fig. 1a and b), is crucial for illustrating a reasonable band structure, unless the band gap closes because of ferroelectricity over the whole structure.\cite{36} The binding energy (BE) with graphene for PbI¬2-termination is 20.5 $meV$/atom, 3.3 times larger than CsI-termination (Table 1).

The electronic band structure of PbI$_2$-(MAI-)terminated Gr/CsPbI$_3$(001)/Gr with PBE+TS+SOC (Fig. 2) reveal some of interesting features. Due to sizable Rashba interactions, in both conduction band (CB) and valence band (VB), surface bands split by momentum $\delta k$ and energy $\delta E$. For the bulk cubic $Pm3m$ (centrosymmetric) CsPbI$_3$ crystal, strong SOC splits the conduction band into one j = 1/2 doublet and one j = 3/2 quartet. Because the valence band is s-like, there is no effective splitting in the highest valence band.\cite{37} However, at the Gr/CsPbI$_3$ interface, we note that s-like valence band at M experiences an asymmetric field with respect to the xy plane and its eigenstate is not s = 1/2 but j = 1/2 being mixed with pz state and other states of adjacent layers. This is manifested in the calculated band structure with non-vanishing Rashba splitting of the surface valence band. In CB, the energy splitting between $j_z$=-1/2 and $j_z$=1/2 is significant, $\delta E$(PbI$_2$-termination) ~ 280 $meV$; this large barrier would hinder electrons to overcome the barrier from the CB extreme in order to directly recombine with the VB extreme.\cite{38} Effective Hamiltonian (eq. (1)) should preserve j=1/2 so that the eigenstate of Rashba split bands for both CB and VB are spanned by $\ket{j=1/2,j_z=\pm 1/2}$. Diagonalization of the Hamiltonian gives an entangled spin-orbital texture which resembles the surface state of topological insulator Bi$_2$Se$_3$ (Fig. S1), “spin-momentum locking”.\cite{39} Indeed, a recent experiment confirmed an emergence of spin-orbital chiral nature by observing circularly polarized light.\cite{40} An explicit calculation of direct transition amplitude from $\ket{j=1/2}$ conduction band to all valence band states $|p|^2 = \sum_{(i=x,y,z)} |p_i|^2$   (Fig. S2a) shows that a direct band-to-band transition is largely suppressed due to spin-orbit entanglement and 2D confinement of wavefunction with the ferroelectric distortion,. We also find that Rashba split band promotes the density of states (Fig. 2c).

\begin{figure}
\centering
\includegraphics[width=8.6cm]{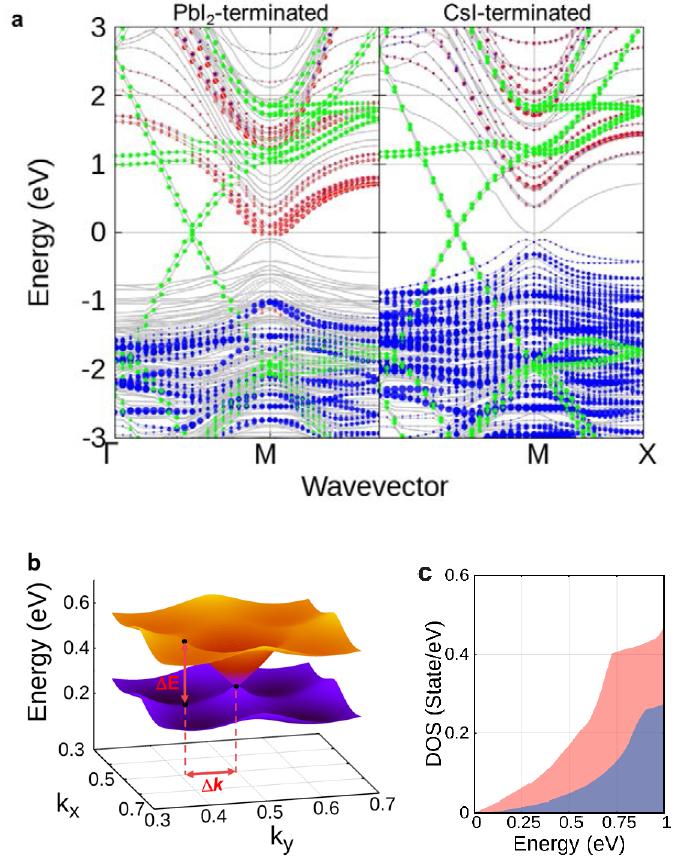}
\caption{\label{Figure. 2} (a) Electronic band structure of 10(9) layers of PbI$_2$-terminated (CsI-terminated) Gr/CsPbI3(001)/Gr-$(1 \times 1)$ symmetric slab with the contributions from the topmost Pb (red), I (blue) and graphene (green). (b) Magnified 2D band of conduction band $\ket{j=1/2,j_z= \pm 1/2}$ around Kramer point M of BZ with the energy difference ($\Delta E$) between the minimum and the upper band and the momentum change ($\Delta k$) between the minimum and M. (c) Density of states in the vicinity of conduction band extremum for Pb p of bulk CsPbI$_3$ (blue) and topmost Pb p (red) of Gr/CsPbI$_3$(001)/Gr.}
\end{figure}

In Gr/CsPbI$_3$(001)/Gr, a surface or gap state shows a rather peculiar structure than usual semi-conductors in which gap states are mainly composed of surface states. For PbI$_2$-termination, while the lowest unoccupied surface state is at the CBM, the highest occupied surface state sits at 1 $eV$ below the VBM. Interestingly, the maximum of VB is composed of bulk state without Rashba splitting due to its centrosymmetry. The surface state shows the opposite trend in CsI-termination. The highest occupied surface state is the VBM, but the lowest unoccupied surface state is 0.5 $eV$ above the CBM. The origin of peculiar energy levels of Gr/CsPbI$_3$(001) can be explained by observing the ferroelectric displacement on CsPbI3 surface. It is found that both CsI- and PbI$_2$-terminated CsPbI$_3$(001) experience an intrinsic ferroelectric displacement resembling the relaxed hetero-interface with graphene (Fig. S3). Compared with unrelaxed and relaxed CsPbI$_3$ slabs, significant energy shifts of surface states are observed (Fig. S4). This natural ferroelectric surface distortion would hint the origin of recent observations of Rashba splitting in CsPbBr$_3$ nanocrystal.\cite{22} However, the role of graphene differs at each termination. At PbI$_2$ termination, graphene further promotes the ferroelectric distortion and the resulting $\alpha_{RD}$ is enhanced to 0.42(VB) and 1.17(CB) compared with 0.18(VB) and 1.00(CB) of the pristine slab. At CsI termination, graphene suppresses the ferroelectric distortion (Fig. 1b and Fig. S3b) and the RD effect is comparable or even less (Table 1). 

The Rashba effect has been shown to exist in bulk MAPbI$_3$: the organic cation MA$^+$ breaks the inversion symmetry and distorts the PbI$_3$ sublattice.\cite{26,27} However, this effect could be local due to orientational disorder of MA+ cations which significantly reduces the Rashba interaction parameter ($\alpha_{RD}$) at the length scale of ~ 3 $nm$. \cite{25,26} Previous work for pristine tetragonal MAPbI$_3$ slab has shown that surface reconstruction could lead to a large RD effect and the effect is more pronounced at PbI$_2$-termination.\cite{41}  We calculated the band structure of rutile TiO$_2$/MAPbI$_3$ (001) interface for PbI$_2$-(MAI-) termination (Fig. 3a and 3b) and also the pristine cubic MAPbI3 slab for comparison (Fig. S5). Despite that the bare rutile (001) is not stable, this facet is favored in device configuration.\cite{42} As reported for tetragonal MAPbI3 slab, $\alpha_{RD}$ of CB and VB in the cubic MAPbI$_3$ slab scales by a factor of 2 (Table 1). Also our result is consistent with the previous work showing larger $\alpha_{RD}$ for PbI$_2$-terminated slab evidenced by the distortions of Pb-I bonds along [001] ($\theta_{Pb-I-Pb}$ ~ 155.5\degree(PbI$_2$-termination), $\theta_{I-Pb-I}$ ~ 170.4 \degree(MAI-termination)) (Fig. S5).\cite{25}

\begin{figure}
\centering
\includegraphics[width=8.6cm]{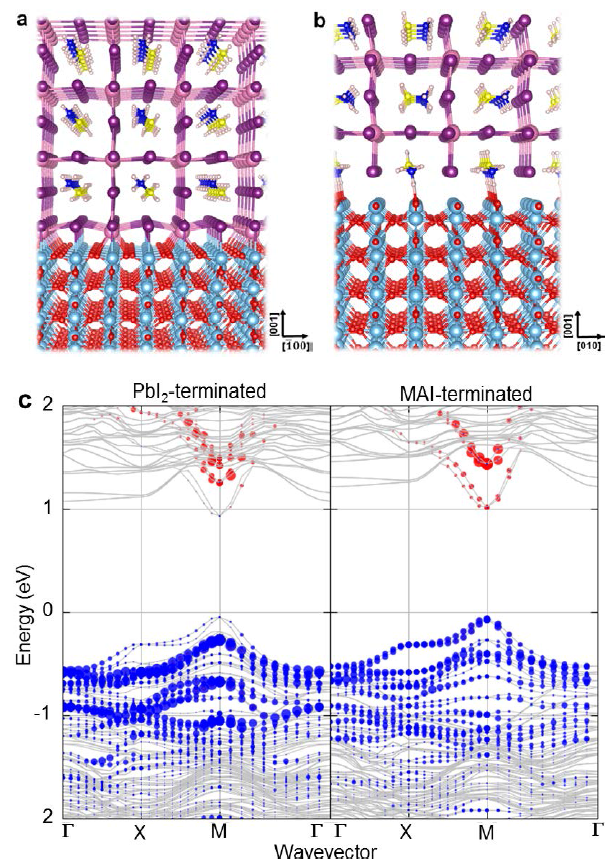}
\caption{\label{Figure. 3} The schematics of optimized (a) PbI$_2$-terminated and (b) MAI-terminated TiO$_2$/cubic MAPbI$_3$-(001) interface. Presence of PbO and TiI bonding in PbI$_2$ termination and SSHB in MAI-termination are highlighted by connecting lines of Pb(light red)-O(red) atoms (~  2.44 $\AA$), Ti(light blue)-I(purple) atoms (~ 2.90 $\AA$) and H(light blue)-O(red) atoms (~ 1.47 $\AA$). (c) Band structure for optimized TiO$_2$/MAPbI$_3$(001) interface calculated with PBE+TS+SOC. The blue(red) circles indicate the contribution of the surface I(Pb) interfaced to TiO$_2$ layer. }
\end{figure}

At the PbI$_2$-termination,\cite{41} we observe the bonding between PbO$_2$ ($d \sim$ 2.44 $\AA$) and TiI ($d \sim$ 2.9 $\AA$).\cite{41} Regarding MAI-terminated interface, our previous work has shown that the orientational freedom of MA$^+$ is significantly curtailed at the TiO$_2$/MAPbI$_3$ interface due to the presence of SSHB between H+ of MA$^+$ and O$^-$ of TiO$_2$.\cite{43} Indeed, AIMD simulations indicate that the O-H bond at the interface remains intact even at 330 K within cubic phase. Moreover, the optimized MA+ orientation at the interface distorts the PbI3 sublattice by increasing the bond distance between Pb and apical I by ~ 0.15 $\AA$. Since SSHB considerably reduces the orientational freedom of MA$^+$ at the interface, the Rashba effect at TiO$_2$/MAPbI$_3$  interface should be more robust than the bulk where it is found to be local or semi-local.\cite{44} Moreover, the near-pinning effect of MA-orientation at the interface was also confirmed by previous DFT/AIMD simulation studies on stable anatase (101) and (001) interfaces with PSC.\cite{32,45} Therefore, the robust RD effect by SSHB is expected to be universal over various TiO2/MAPbI3 interface configurations.

In general, the VBM/CBM related states of TiO$_2$ and MAPbI$_3$ lie at high symmetry $\Gamma$ and M of Brillouin zone (BZ), respectively. The energy splitting is considerably larger for CB (Fig. 3c). This is anticipated as MAPbI$_3$ CB is dominated by heavier Pb states compared with valence states of relatively lighter I atoms. On the other hand, we note that the RD effect has been suppressed at PbI2-terminated interface because of stabilization of the surface via PbO$_2$ and TiI bonding with BE ~ 15.6 $meV$/atom. We also observe that a significant distortion along [001] has been quenched ($\theta_{Pb-I-Pb}$ ~ 172.1\degree). Therefore, $\alpha_{RD}$ ~ 0.08 (VB) and 0.17 (CB) in TiO$_2$-termination interface is significantly reduced compared to the pristine slab (Table 1). In MAI-terminated interface, SSHB results in enhancing the RD effect for $\alpha_{RD}$ being ~ 0.29 (VBM) and 0.58 (CBM). A momentum (k-space) mismatch is also found between the VBM and CBM as indicated by different Δk for CBM and VBM which suppresses the recombination and increases the carrier lifetime.\cite{46} As in the case of Gr/CsPbI$_3$, a direct band-to-band recombination of MAPbI3 is significantly quenched being interfaced to TiO$_2$. (Fig. S2b) All the calculated RD interaction parameters $\alpha_{RD}$ are in Table 1.

\begin{figure}
\centering
\includegraphics[width=8.6cm]{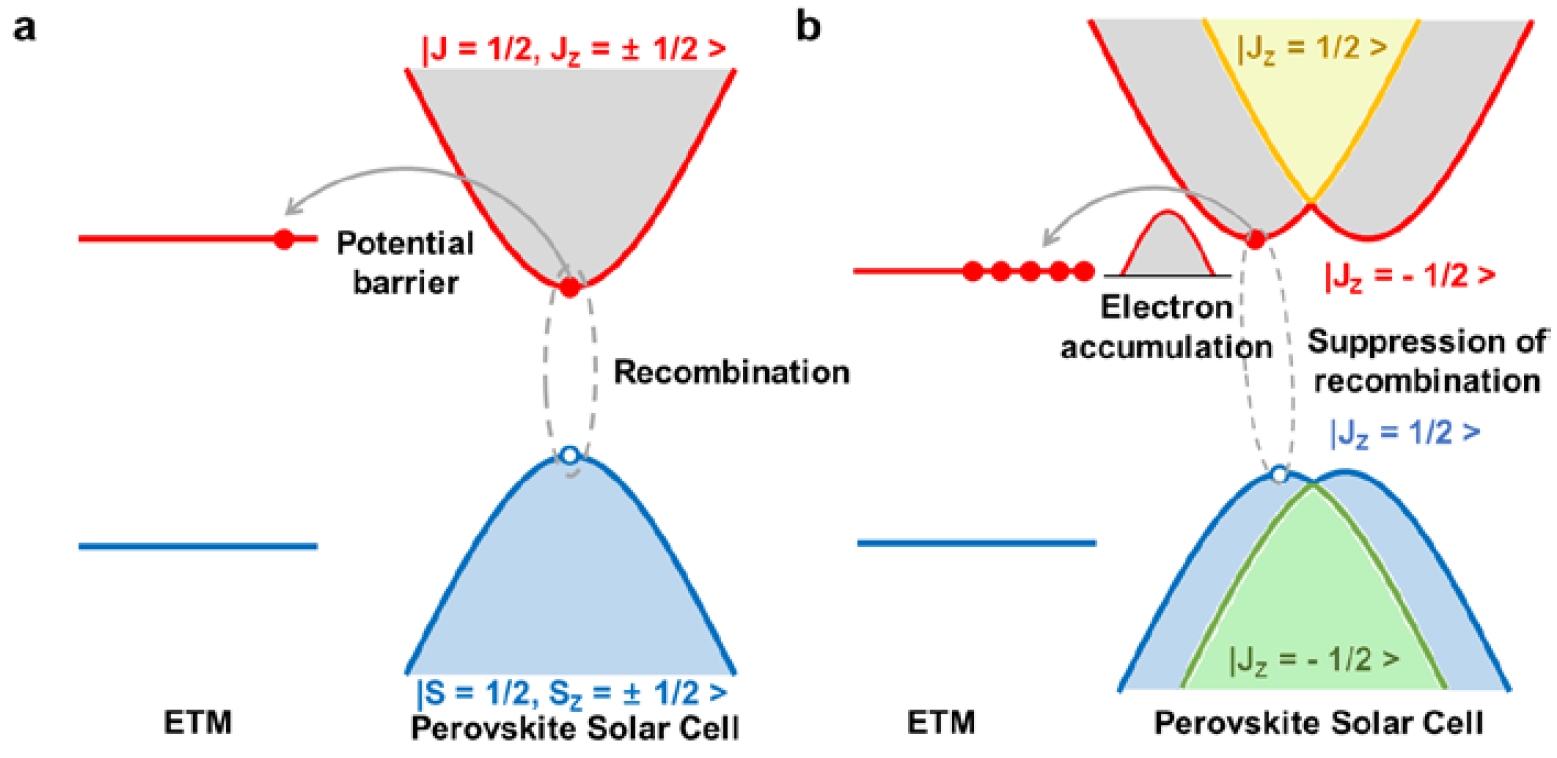}
\caption{\label{Figure. 4} Schematic of electron transfer process with and without Rashba-Dresselhaus Effect at the interface between perovskite solar cell and electron transfer material. (a) Without the RD effect, the large potential barrier between PSC and ETM cannot be overcome by a rapid electron(red circle)-hole(red circle) pair recombination process. (b) With the RD effect, long lifetime of the electron-hole pair contributes to the accumulation of electrons to overcome the barrier.}
\end{figure}

The direct optical measurement and photoemission study have shown that a sizeable electron transfer barrier of ~ 0.4 eV may exist at TiO$_2$/MAPbI$_3$ interface.\cite{47,48} In the presence of such a large barrier, electron transfer process should be strongly impeded, resulting in accumulation of electrons at interface. In the case that the charge recombination process is favorable, this should result in considerable degradation in device performance. However, the presence of Rashba splitting at the interface (Fig. 3 and Table 1) reduces such recombination process until the potential barrier is overcome by electron accumulation (Fig. 4). Therefore, the presence of Rashba effect at the interface is critical for improving the electron transfer process at the TiO$_2$/MAPbI$_3$ interface. 

Apart from 0 K DFT result, to elucidate the impact of thermal degrees of freedom, we average $\alpha_{RD}$ from NVT ensemble based on AIMD simulations around 600 K (for cubic CsPbI$_3$)\cite{49}  and 330 K (for cubic MAPbI3).\cite{50}  We choose the most pronounced case, PbI$_2$-termination for Gr/CsPbI$_3$/Gr and MAI-termination TiO$_2$/MAPbI$_3$. Initially we assumed that the RD effect is overestimated in the 0 K DFT. However, the thermal average of RD splitting is found to be comparable to 0 K DFT result (Table 1). Thermal average of $\alpha_{RD}$(600 K) for PbI2-terminated Gr/CsPbI$_3$ is 0.46($\pm$0.15) $eV \cdot \AA$ (VB) and 1.25($\pm$0.54) $eV \cdot \AA$ (CB). $\alpha_{RD}$(330 K) for MAI-terminated TiO$_2$/MAPbI$_3$, where the alignment of MA at the TiO$_2$ interface is kept fixed, is 0.26($\pm$0.07) eV·Å (VB) and 0.61($\pm$0.15) eV·Å (CB). In particular, the surface I’s configuration does not stray from 0 K configuration even at 330 K because of the enrooted MA (Fig. 3b). Therefore, VB has almost the same $\alpha_{RD}$ as that of 0 K DFT with a negligible standard deviation ~ 0.07$eV \cdot \AA$.

In summary, we have shown two examples of inorganic/organic perovskite solar cell interfaces: Gr/CsPbI$_3$, a promising interface for efficient carrier transport and TiO2/MAPbI3, a common electron transport layer interface. We report the sizable Rashba interaction $\alpha_{RD}$(0 K) ~ 1.17 (0.42) $eV \cdot \AA$ and $\alpha_{RD}$(600 K) ~ 1.25 (0.46) $eV \cdot \AA$ for electron (hole) at PbI2-terminated Gr/CsPbI$_3$. The TiO$_2$/MAPbI$_3$ interface also shows a significant RD interaction $\alpha_{RD}$(0 K) ~ 0.58 (0.29) $eV \cdot \AA$ and $\alpha_{RD}$(330 K) ~ 0.61 (0.26) $eV \cdot \AA$ for electron (hole) carriers. The enrooted alignment of MA+ at the TiO$_2$ interface gives rise to a strong and firm RD effect even at high temperatures (> 330 K) with small variance. Because of large SOC nature and geometrical complexity of PSCs, its interface with other layers poses rich phenomena. A clever manipulation of such interfaces could accelerate further improvement of the PSC efficiency.

\begin{table}
\caption{ Binding energy (BE) and Rashba-Dresselhaus parameter $\alpha_{RD}$ for graphene-, TiO$_2$-interfaced cubic perovskite solar cell materials along BZ depending on the terminations, PbI$_2$ and AI where A refers to the A-site cation, Cs$^+$ or MA$^+$.}
    \begin{tabular}{llll}
    
    \toprule
    \hline
    \multicolumn{1}{c}{Interface} & \multicolumn{1}{c}{BE (meV/atom)} & \multicolumn{2}{c}{$\alpha_{RD}$} \\ \hline
    \midrule
                                  &                                  & CB                       & VB       \\ \hline
     Gr/CsPbI$_3$/Gr (PbI$_2$)    & 20.54   & 0.42 & 1.17   \\ \hline
     Pristine CsPbI$_3$ (PbI$_2$) & -       & 0.18 & 1.00   \\ \hline
     Gr/CsPbI$_3$/Gr (CsI)        & 6.23    & 0.59 & 0.50   \\ \hline
     Pristine CsPbI$_3$ (CsI)     & -       & 0.53 & 0.54   \\ \hline
     TiO$_2$/MAPbI$_3$ (PbI$_2$)  & 15.64   & 0.08 & 0.17   \\ \hline
     Pristine MAPbI$_3$ (PbI$_2$) & -       & 0.78 & 0.81   \\ \hline
     TiO$_2$/MAPbI$_3$ (MAI)      & 10.33   & 0.29 & 0.58   \\ \hline
     Pristine MAPbI$_3$ (MAI)     & -       & 0.18 & 0.30   \\
    \hline
    \bottomrule
    \end{tabular}
\end{table}

We used Vienna Ab initio Simulation Package (VASP)\cite{51} for non-collinear DFT calculations using PBE functional plus Tkatchenko-Scheffler(TS)\cite{52}/Grimme DFT-D3\cite{53}  van der Waals correction with inclusion of spin-orbit coupling by switching off any presumed symmetry. Our previous work has shown that GGA+SOC results are consistent with that of higher level but computationally expensive HSE06\cite{54} +SOC calculations.\cite{36} For Gr/CsPbI$_3$(001) system, we used $(4 \times 4 \times 1)$ kmesh for sampling the Brillouin zone and 500 eV for the energy cutoff. As we checked the convergence of band gap with respect to the thickness of CsPbI$_3$ slab, the convergence has been met from 6 cubic CsPbI$_3$ layers. For TiO$_2$/MAPbI$_3$(001) system, we used $(6 \times 6 \times 1)$ kmesh with 520 eV energy cutoff. A supercell consists of 11 rutile TiO$_2$ layers and 3 cubic MAPbI$_3$ layers with (001) orientation for both, where the lattice mismatch using TiO$_2$(001)-$(\sqrt(2) \times \sqrt(2) \times 1)$ is as small as ~ 3 $\%$. A vacuum size of ~ 30 $\AA$ is included. We also used Quantum ESPRESSO package v.6.1\cite{55} with fully relativistic PAW PBE pseudopotential for Pb 6p6s5d, I 5p5s and Cs 6s5p5s at the energy cutoff of 40 Ry. Ab initio MD simulations with time step $\delta t$ = 0.5 $fs$ were performed with total duration of 12 $ps$ and 30 $ps$ at 600 K and 330 K for Gr/CsPbI$_3$ and TiO$_2$/MAPbI$_3$, respectively. We sampled NVT configuration using Nosé thermostat and discarded 3 $ps$ for the initialization.

\begin{acknowledgments}
C.W.M. conceived the idea, performed DFT and AIMD simulations and analyzed the data. S.J. helped in DFT calculation. All discussed and C.W.M., K.S.K. and G.L. wrote the manuscript. This work was supported by National Honor Scientist Program (2010-0020414) and Basic Science Research Program (2015R1C1A1A01055922) of NRF. Computation was supported by KISTI (KSC-2017-S1-0025, KSC-2017-C3-0081).
\end{acknowledgments}

\bibliographystyle{apsrev4-1}
\bibliography{myungbib}

\end{document}